\documentclass[amsmath,amssymb,floatfix,prb,superscriptaddress,aps]{revtex4}

\usepackage{graphicx}
\usepackage{epsfig}
\usepackage{dcolumn}
\usepackage{bm}
\usepackage{times}
\begin{document}


\title{Electron dynamics inside a vacuum tube diode through linear differential equations}
\author{Gabriel Gonz\'alez}
\email{gabrielglez@iteso.mx}
\author{Fco. Javier Gonz\'alez Orozco}
\affiliation{Departamento de Matem\'aticas y F\'isica, Instituto Tecnol\'ogico y de Estudios Superiores de Occidente, \\ Perif\'erico Sur Manuel G\'omez Mor{\'i}n 8585 C.P. 45604, Tlaquepaque, Jal., MEXICO}

\date{\today}
\begin{abstract}
In this paper we analyze the motion of charged particles in a vacuum tube diode by solving linear differential equations. Our analysis is based on expressing the volume charge density as a function of the current density and coordinates only, i.e. $\rho=\rho(J,z)$, while in the usual scheme the volume charge density is expressed as a function of the current density and electrostatic potential, i.e. $\rho=\rho(J,V)$. Our approach gives the well known behavior of the classical current density proportional to the three-halves power of the bias potential and
inversely proportional to the square of the gap distance between the electrodes, and does not require the solution of the nonlinear differential equation normally associated with the Child-–Langmuir formulation.
\end{abstract}
\maketitle

\section{Introduction}
The motion of charged particles accelerated across a
gap is of wide interest in fields such as high power
diodes and vacuum microelectronics. Child and Langmuir first
studied the space charge limited emission for two infinite parallel plane electrodes
at fixed voltage $V_0$ in vacuum separated by a distance $D$.\cite{child,lang} The charges produced at the cathode are
made to increase (e.g. by increasing the temperature in
a thermionic diode or by increasing the power of a laser
in case of a photocathode). Interestingly, a saturation is
observed and the current is said
to be space charge limited. A useful approximation for
the amount of current flowing in such cases is the Child-Langmuir expression for transmitted current. For electrodes having a potential
difference $V$ and separated by a distance $D$, the Child-Langmuir law is obtained by solving Poisson's equation
\begin{equation}
\frac{d^2V}{dz^2}=-\frac{\rho}{\epsilon_0}
\label{eq1}
\end{equation}
where $V$ is the electrostatic potential, $\rho$ is the volume charge density and $\epsilon_0$ is the permittivity of free space.\cite{poll} Because this diode is assumed to have infinite extension in the x-y directions, we can define the current density by
\begin{equation}
J(z)=\rho(z)v(z)=-J_{CL}
\label{eq2}
\end{equation}
where $v$ is the velocity of the electrons. By charge conservation the current density can not vary with $z$, hence the current density is constant. Now we can find the velocity of the electrons by conservation of energy
\begin{equation}
\frac{mv^2}{2}-eV=0
\label{eq3}
\end{equation}  
where $m$ and $e$ are the electron's mass and charge, respectively. In Eq.(\ref{eq3}) we have assumed that the electron is initially at rest in the grounded cathode. Solving Eq.(\ref{eq3}) for the velocity and substituting in Eq.(\ref{eq2}) we obtain the volume charge density as a function of the current density and the electrostatic potential
\begin{equation}
\rho(z)=-\frac{J_{CL}}{\sqrt{2eV/m}}
\label{eq4}
\end{equation}
Substituting Eq.(\ref{eq4}) into Eq.(\ref{eq1}) we have a second order nonlinear differential equation for the electrostatic potential
\begin{equation}
\frac{d^2V}{dz^2}=\frac{J_{CL}}{\epsilon_0\sqrt{2eV/m}}
\label{eq5}
\end{equation}
with the following boundary conditions
\begin{equation}
\frac{dV}{dz}\Bigg|_{z=0}=0 \quad \mbox{and} \quad V(z)\Bigg|_{z=0}=0
\label{eq6}
\end{equation}
Using the following ansatz for Eq.(\ref{eq5})
\begin{equation}
V(z)=V_0\left(\frac{z}{D}\right)^p
\label{eq6a}
\end{equation}
and substituting Eq.(\ref{eq6a}) into Eq.(\ref{eq5}) we find that the non linear differential equation is satisfied for 
\begin{equation}
p=\frac{4}{3}\quad \mbox{and}\quad J_{CL}=\frac{4\epsilon_0}{9D^2}\sqrt{\frac{2e}{m}}V_0^{3/2}
\label{eq6b}
\end{equation}
The volume charge density in the gap is determined by Eq.(\ref{eq4})
\begin{equation}
\rho(z)=-\frac{4\epsilon_0V_0}{9D^2}\left(\frac{D}{z}\right)^{2/3}
\label{eq6c}
\end{equation} 
The main result is the Child-Langmuir law which states that the behavior of the current density is proportional to the three-halves power of the bias potential and inversely proportional to the square of the gap distance between the electrodes.\\
Since the derivation of this fundamental law many important and useful variations on the classical Child-Langmuir law have been investigated to account for special geometries,\cite{lang1,lang2,page} relativistic electron energies,\cite{jory} non zero initial electron velocities,\cite{lang3,jaffe} quantum mechanical effects,\cite{lau,ang} nonzero electric field at the cathode surface,\cite{barbour} and using vacuum capacitance.\cite{um}\\
Given the wide interest and applicability in this topic we present the analysis of the motion of charged particles in a vacuum tube diode by solving linear differential equations. Our derivation is based on expressing the volume charge density as a function of the current density and coordinates only, i.e. $\rho=\rho(J,x)$. Our approach gives the well known behavior of the classical current density proportional to the three-halves power of the bias potential and
inversely proportional to the square of the gap distance between the electrodes, and does not require the solution of the nonlinear differential equation normally associated with the Child-–Langmuir formulation.
\section{Alternative approach}
\begin{figure}[!ht]
  \begin{center}
    \begin{tabular}{cc}
      \resizebox{52mm}{!}{\includegraphics{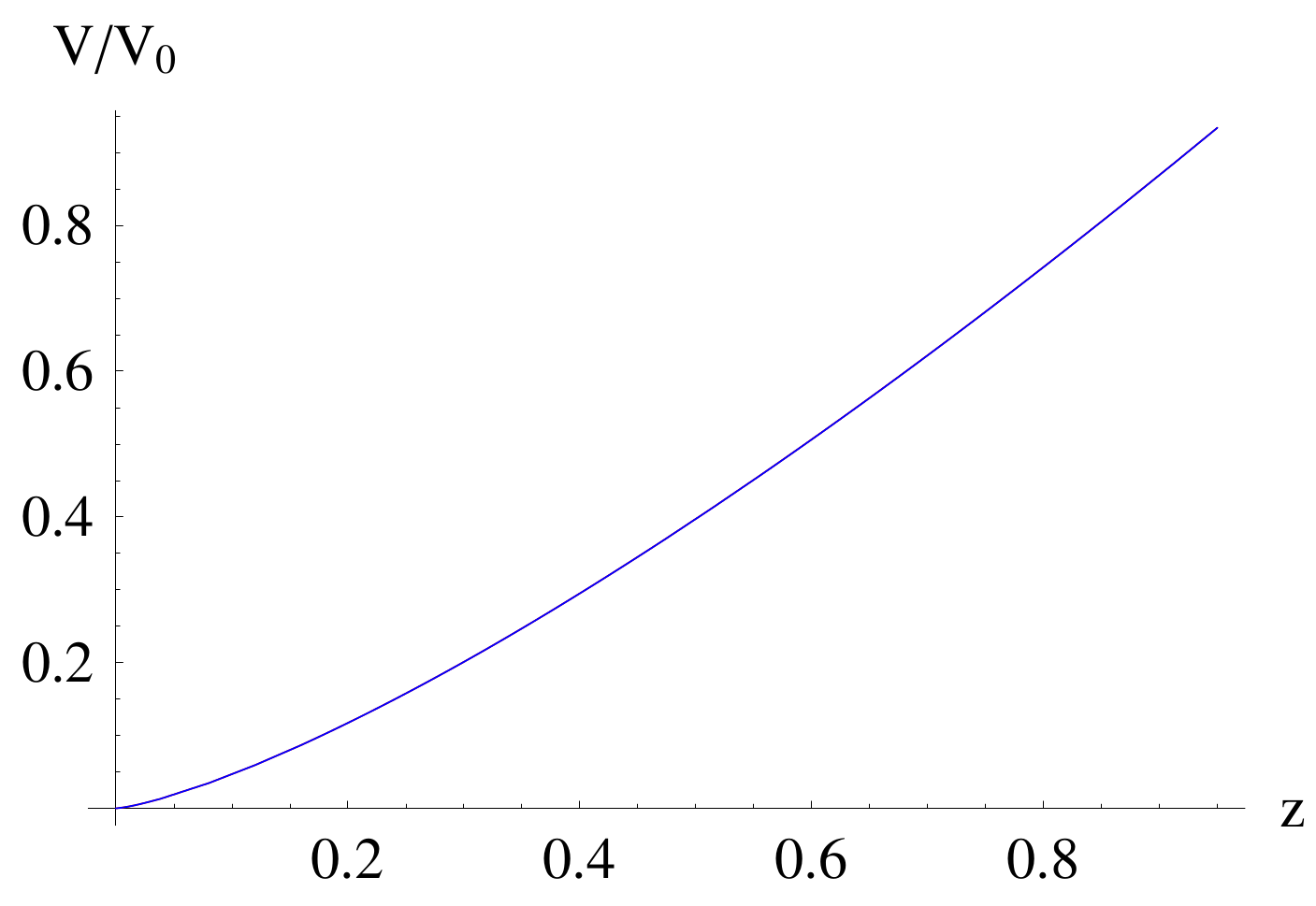}} &
      \resizebox{52mm}{!}{\includegraphics{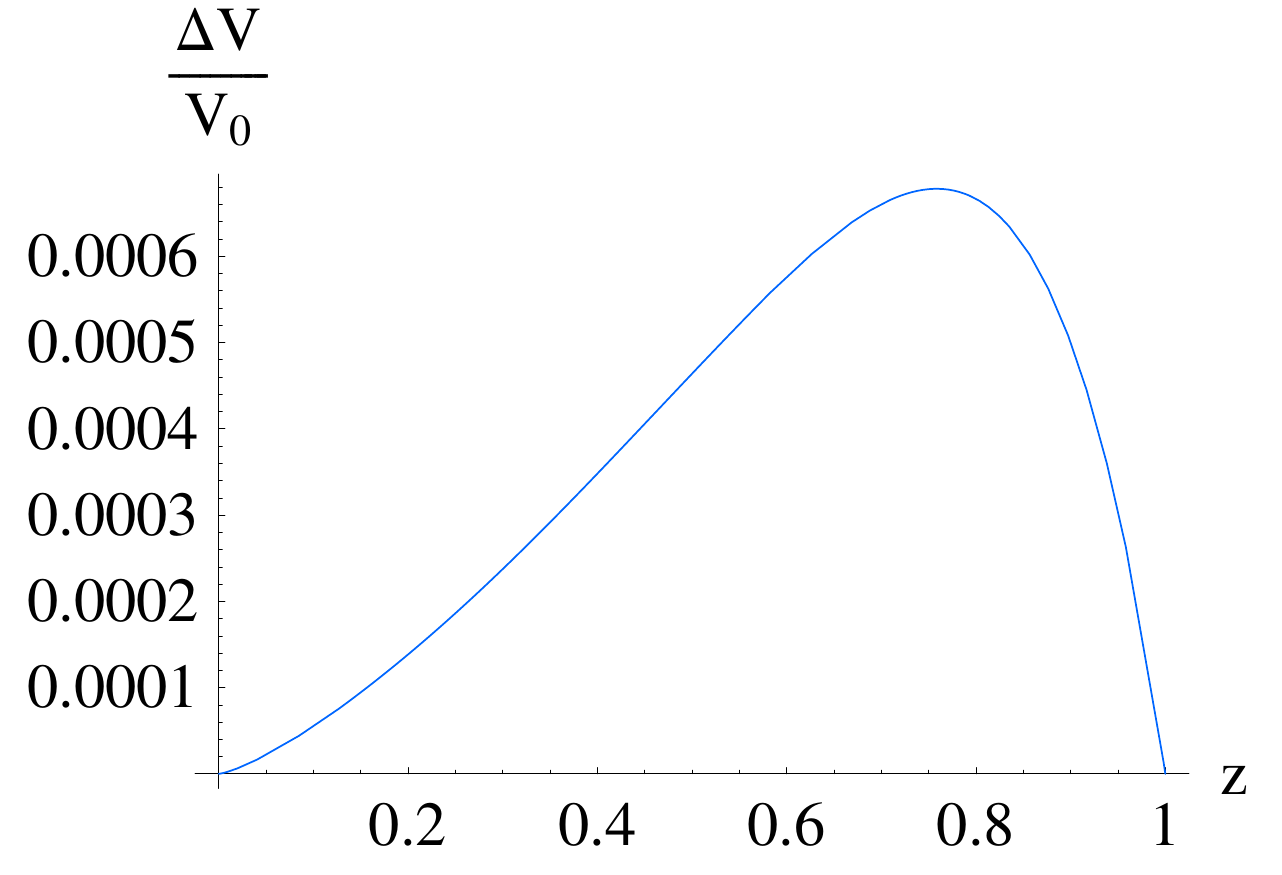}} \\ 
    \multicolumn{1}{c}{\mbox{\bf (a)}} &
		\multicolumn{1}{c}{\mbox{\bf (b)}} 
    \end{tabular}
    \caption{Plots showing (a) the electrostatic potential in the adiabatic approximation with $D=1$, (b) the difference between the electrostatic potentials in the adiabatic and non adiabatic approximation. Note how the functions for the electrostatic potential have the same value at $z=0$ and $z=D$ such that the applied bias in the vacuum tube diode is $V_0$ and are really close to each other for $0<z<D$.}
	\label{potencial}
  \end{center}
\end{figure}

\begin{figure}[ht]
\centering
\includegraphics[scale =0.5] {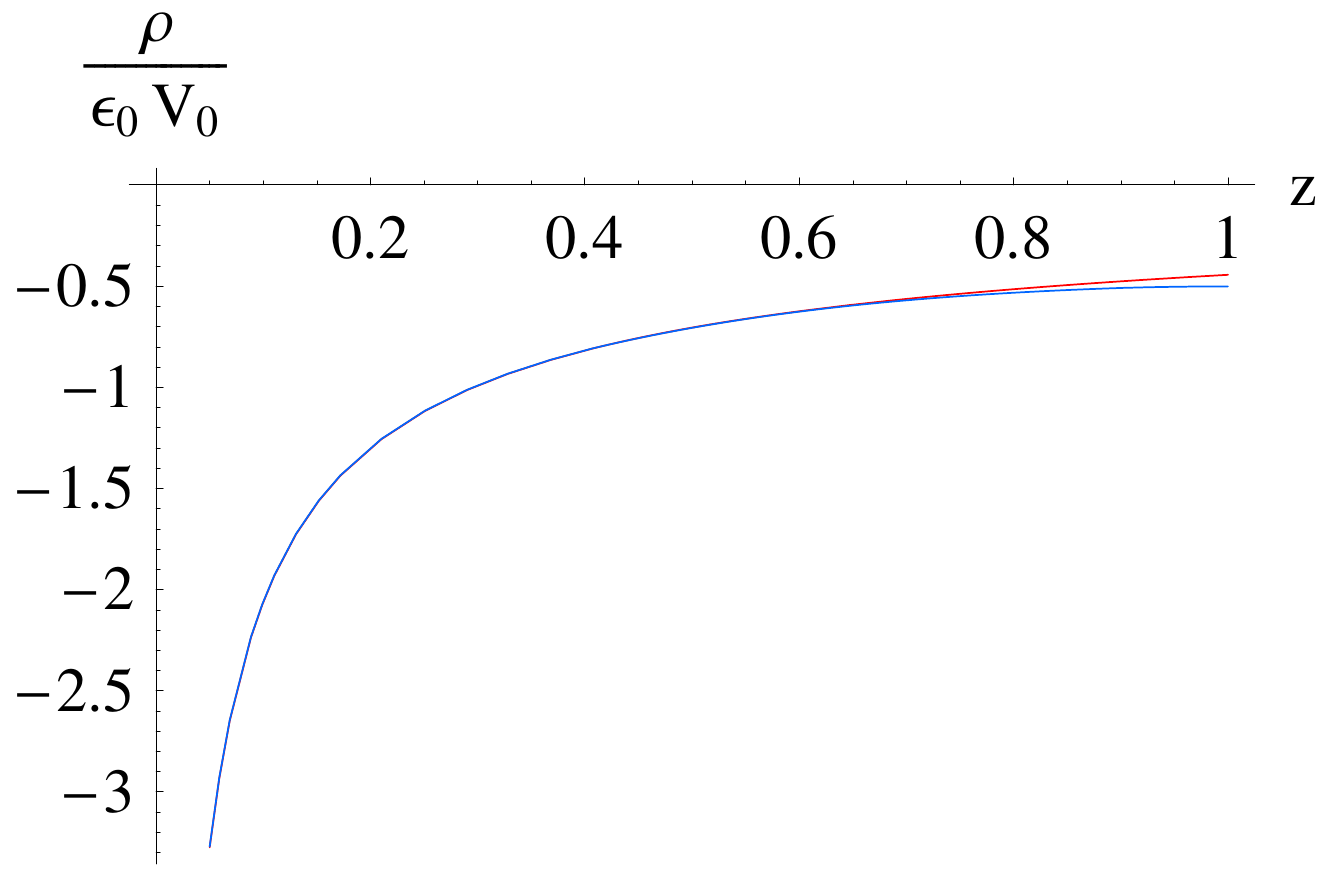}
\label{carga}
\caption{We have plotted the graphs of the volume charge density for the adiabatic and non adiabatic regime with $D=1$. Note how the functions superimposed except at $z=D$ where the volume charge density for the non adiabatic regime has an horizontal slope.}
\end{figure}

In the Child-Langmuir derivation we have expressed the volume charge density as a function of the current density and the electrostatic potential, let us now try to express the volume charge density as a function of the current density and the spatial coordinate. The velocity of a charged particle subjected to an electric field is governed by Newton's second law 
\begin{equation}
m\frac{dv}{dt}=-eE(z)
\label{eq7}
\end{equation}
where $e=1.6\times 10^{-19}$C and $E(z)$ is the electric field. We can obtain the electric field inside the gap by integrating Gauss's law from $z=0$ to an arbitrary position in the gap
\begin{equation}
E(z)=\frac{1}{\epsilon_0}\int_{0}^{z}\rho(z)dz
\label{eq8}
\end{equation} 
where we have used the fact that $E(z=0)=0$. Integrating by parts Eq.(\ref{eq8}) we have
\begin{equation}
E(z)=\frac{\rho}{\epsilon_0}z-\frac{1}{\epsilon_0}\int_{0}^{z}z\frac{d\rho}{dz}dz
\label{eq9}
\end{equation}
Let us work now in the adiabatic regime and neglect the contribution of $\rho^{\prime}(z)$ in Eq.(\ref{eq9}), i.e. the volume charge density is slowly varying in space. Substituting Eq.(\ref{eq9}) into Eq.(\ref{eq7}) we have
\begin{equation}
m\frac{dv}{dt}=\frac{d}{dz}\left(\frac{mv^2}{2}\right)=-\frac{e\rho}{\epsilon_0}z
\label{eq10}
\end{equation}
Integrating Eq.(\ref{eq10}) from $z=0$ to an arbitrary position we have
\begin{equation}
\frac{mv^2}{2}=-\frac{e\rho}{2\epsilon_0}z^2+\frac{e}{2\epsilon_0}\int_{0}^{z}z^2\frac{d\rho}{dz}dz
\label{eq11}
\end{equation}
where we have integrated by parts the right hand term. Applying the adiabatic condition to Eq.(\ref{eq11}) we obtain
\begin{equation}
\frac{mv^2}{2}=-\frac{e\rho}{2\epsilon_0}z^2
\label{eq12}
\end{equation}
Solving for the velocity in Eq.(\ref{eq2}) and substituting into Eq.(\ref{eq12}) we have
\begin{equation}
\frac{mJ^2}{2\rho^2}=-\frac{e\rho}{2\epsilon_0}z^2
\label{eq13}
\end{equation} 
Using Eq.(\ref{eq13}) we can express the volume charge density as a function of the current density and coordinates as
\begin{equation}
\rho(z)=-\left(\frac{mJ^2\epsilon_0}{ez^2}\right)^{1/3}
\label{eq14}
\end{equation}
Substituting Eq.(\ref{eq14}) into Eq.(\ref{eq1}) we have 
\begin{equation}
\frac{d^2V}{dz^2}=\frac{1}{\epsilon_0}\left(\frac{mJ^2\epsilon_0}{ez^2}\right)^{1/3}
\label{eq15}
\end{equation}
Equation (\ref{eq15}) is a linear second order diferential equation which can be solved by quadratures. Integrating Eq.(\ref{eq15}) twice and using the initial conditions given in Eq.(\ref{eq6}) we obtain the solution for the electrostatic potential
\begin{equation}
V(z)=\frac{9}{4\epsilon_0}\left(\frac{mJ^2\epsilon_0z^4}{e}\right)^{1/3}
\label{eq16}
\end{equation}
If we apply the boundary condition of the fixed potential anode $V(z=D)=V_0$ in Eq.(\ref{eq16}), we can find the current density in the adiabatic regime which is given by
\begin{equation}
|J|=\frac{8\epsilon_0}{27D^2}\sqrt{\frac{e}{m}}V_0^{3/2}
\label{eq16a}
\end{equation}
Equation (\ref{eq16a}) is remarkably close to the result for $J_{CL}$ derived in Eq.(\ref{eq6b}), i.e. $|J|\approx0.47J_{CL}$.
Substituting Eq.(\ref{eq16a}) into equations (\ref{eq14}) and (\ref{eq16}) we find
\begin{equation}
\rho(z)=-\frac{4\epsilon_0V_0}{9D^2}\left(\frac{D}{z}\right)^{2/3} \quad \mbox{and}\quad V(z)=V_0\left(\frac{z}{D}\right)^{4/3}
\label{eq16b}
\end{equation}
Equations (\ref{eq16b}) are the same exact results derived with the Child-Langmuir formulation.  \\ 
If we want to go beyond the adiabatic approximation, we integrate again by parts Eq.(\ref{eq9}) to obtain
\begin{equation}
E(z)=\frac{\rho}{\epsilon_0}z-\frac{\rho^{\prime}}{2\epsilon_0}z^2+\frac{1}{2\epsilon_0}\int_{0}^{z}z^2\frac{d^2\rho}{dz^2}dz
\label{eq17}
\end{equation}
Assuming now that we can neglect the last term in Eq.(\ref{eq17}) we have in Newton's second law
\begin{equation}
m\frac{dv}{dt}=\frac{d}{dz}\left(\frac{mv^2}{2}\right)=-\frac{e\rho}{\epsilon_0}z+\frac{e\rho^{\prime}}{2\epsilon_0}z^2
\label{eq18}
\end{equation}
Integrating by parts Eq.(\ref{eq18}) we have
\begin{equation}
\frac{mv^2}{2}=-\frac{e\rho}{2\epsilon_0}z^2+\frac{e}{3\epsilon_0}z^3\frac{d\rho}{dz}-\frac{e}{3\epsilon_0}\int_{0}^{z}z^3\frac{d^2\rho}{dz^2}dz
\label{eq19}
\end{equation}
Neglecting the last term in Eq.(\ref{eq19}) we have
\begin{equation}
\frac{mv^2}{2}=-\frac{e\rho}{2\epsilon_0}z^2+\frac{e}{3\epsilon_0}z^3\frac{d\rho}{dz}
\label{eq20}
\end{equation}
Solving for the velocity in Eq.(\ref{eq2}) and substituting into Eq.(\ref{eq20}) we end up with
\begin{equation}
\frac{mJ^2}{2\rho^2}=-\frac{e\rho}{2\epsilon_0}z^2+\frac{e}{3\epsilon_0}z^3\frac{d\rho}{dz}
\label{eq21}
\end{equation} 
Equation (\ref{eq21}) is a linear differential equation over $\rho^3$. Solving Eq.(\ref{eq21}) for the charge density we find
\begin{equation}
\rho^3(z)=-\frac{9m\epsilon_0J^2}{13ez^2}+Cz^{9/2}
\label{eq22}
\end{equation}
where $C$ is a constant of integration. If we apply the following condition $\rho^{\prime}(z=D)=0$ in Eq.(\ref{eq22}), i.e. the charge density is minimum at the anode, we have the following charge distribution in the vacuum tube diode 
\begin{equation}
\rho(z)=-\left(\frac{m\epsilon_0J^2}{13ez^2}\right)^{1/3}\left[9+4\left(\frac{z}{D}\right)^{13/2}\right]^{1/3}
\label{eq23}
\end{equation} 
Substituting Eq.(\ref{eq23}) into Eq.(\ref{eq1}) we have 
\begin{equation}
\frac{d^2V}{dz^2}=\frac{1}{\epsilon_0}\left(\frac{m\epsilon_0J^2}{13ez^2}\right)^{1/3}\left[9+4\left(\frac{z}{D}\right)^{13/2}\right]^{1/3}
\label{eq24}
\end{equation}
Equation (\ref{eq24}) is a linear differential equation which can be solved by quadratures. Integrating Eq.(\ref{eq24}) with Mathematica we find
\begin{equation}
V(z)=\frac{1}{\epsilon_0}\left(\frac{243m\epsilon_0J^2}{13e}\right)^{1/3}\left\{z^{4/3}\left[_{2}F_1\left(-\frac{1}{3},\frac{2}{39};\frac{41}{39};-\frac{4}{9}\left(\frac{z}{D}\right)^{13/2}\right)-\frac{1}{4} {_{2}}F_1\left(-\frac{1}{3},\frac{8}{39};\frac{47}{39};-\frac{4}{9}\left(\frac{z}{D}\right)^{13/2}\right)\right]\right\}
\label{eq25}
\end{equation}
where $_{2}F_1(a,b;c;z)$ is the Hypergeometric function.\cite{leb} 
If we apply the boundary condition of the fixed potential anode $V(z=D)=V_0$ in Eq.(\ref{eq25}), we can find the current density which is given by
\begin{equation}
|J|=\frac{\epsilon_0}{D^2}\sqrt{\frac{13e}{243m\left[_{2}F_1\left(-\frac{1}{3},\frac{2}{39};\frac{41}{39};-\frac{4}{9}\right)-\frac{1}{4} {_{2}}F_1\left(-\frac{1}{3},\frac{8}{39};\frac{47}{39};-\frac{4}{9}\right)\right]^{3}}}V_0^{3/2}
\label{eq26}
\end{equation}
Evaluating Eq. (\ref{eq26}) numerically we find that $|J|\approx0.57J_{CL}$. 
Our approach can be easily extended to include the next order derivatives in the volume charge density to improve the accuracy of the current charge density, unfortunately this will lead us to solve non linear differential equations to express the volume charge density as a function of the current density and spatial coordinate and the expressions for the electrostatic potential and charge distribution become far more complex.\\
In Fig. (\ref{potencial}) we have plotted the electrostatic potential in the adiabatic regime and we have plotted the difference between the electrostatic potential in the adiabatic and non adiabatic regime, respectively. Note how there is no difference at $z=0$ and $z=D$ for the electrostatic potential function in order for the applied bias to be $V_0$ in the vacuum tube diode.
In Fig. (\ref{carga}) we have plotted the volume charge density for the adiabatic and non adiabatic regime where we can see that the functions are practically identical except at $z=D$ where the volume charge density for the non adiabatic regime has an horizontal slope.

\section{Conclusions}
The alternative method presented in this article of deriving the $V^{3/2}/D^2$ scaling of the Child-Langmuir law avoids the need of solving a nonlinear differential equation and presents a new insight into the way of approaching the problem of the charge dynamics inside a vacuum tube diode. 
In the Child-Langmuir formulation, the Poisson equation is solved with a volume charge distribution that involves the electrostatic potential in contrast with our procedure which provides the volume charge distribution as a function of coordinates, so that the solution can be readily determined by simply integrating Poisson's equation and applying the boundary conditions at the cathode and anode, respectively.
In addition, we have obtained solutions for the electrostatic potential and volume charge density in the adiabatic and non adiabatic approximations. We have shown that one has the same solution for the electrostatic potential and volume charge density as in the Child-Langmuir formulation in the adiabatic approximation but with a different current density.


\begin{thebibliography}{99}
\bibitem{child} C.D. Child, ``Discharge from hot CaO", Phys. Rev. {\bf 32}, 492-511 (1911) 
\bibitem{lang} I. Langmuir, ``The effect of space charge and residual gases on thermionic currents in high vacuum", Phys. Rev. {\bf 2}, 450-486 (1913)
\bibitem{poll} Gerald L. Pollak and Daniel R. Stump, \textit{Electromagnetism},(Addison-Wesley, 2002)
\bibitem{lang1} I. Langmuir and K.B. Blodgett, ``Currents limited by space charge between coaxial cylinders", Phys. Rev. {\bf 22}, 347-356 (1923)
\bibitem{lang2} I. Langmuir and K.B. Blodgett, ``Currents limited by space charge between concentric spheres", Phys. Rev. {\bf 24}, 49-59 (1924)
\bibitem{page} L. Page and N.I. Adams, Jr., ``Space charge between coaxial cylinders", Phys. Rev. {\bf 68}, 126-129 (1945)
\bibitem{jory} H. R. Jory and A.W. Trivelpiece, ``Exact relativistic solution for the one dimensional diode", J. Appl. Phys. {\bf 40}, 3924-3926 (1969) 
\bibitem{lang3} I. Langmuir, ``The effect of space charge and initial velocities on the potential distribution and thermionic current between parallel plane electrodes", Phys. Rev. {\bf 21}, 419-435 (1923)
\bibitem{jaffe} G. Jaff\'e, ``On the currents carried by electrons of uniform initial velocity", Phys. Rev. {\bf 65}, 91-98 (1944)
\bibitem{lau} Y.Y. Lau, D. Chernin, D.G. Colombant, and P.T. Ho, ``Quantum extension of Child-Langmuir law", Phys. Rev. Lett. {\bf 66}, 1446-1449 (1991)
\bibitem{ang} L.K. Ang, T.J.T. Kwan, and Y.Y. Lau, ``New scaling of Child-Langmuir law in the quantum regime", Phys. Rev. Lett. {\bf 91}, 208303-1-208303-4 (2003)
\bibitem{barbour} J.P. Barbour, W.W. Dolan, J.K. Trolan, E.E. Martin, and W.P. Dyke, ``Space-charge effects in field emission", Phys. Rev. {\bf 92}, 45-51 (1953)
\bibitem{um} R.J. Umstattd, C.G. Carr, C.L. Frenzen, J.W. Luginsland, and Y.Y. Lau, ``A simple phyhsical derivation of Child-Langmuir space-charge-limited emission using vacuum capacitance", Am. J. Phys. {\bf 73}, 160-163 (2005)
\bibitem{leb} N.N. Lebedev, {\it Special functions and their applications}, (Dover Publications Inc., New York, 1972)
\end{thebibliography}
\end{document}